# Observation of topological Faraday and Kerr rotations in quantum anomalous Hall state by terahertz magneto-optics

Ken N. Okada[1,2]†*, Youtarou Takahashi[1,2,3]*, Masataka Mogi[2], Ryutaro Yoshimi[2], Atsushi Tsukazaki[3,4], Kei S. Takahashi[1,3], Naoki Ogawa[1], Masashi Kawasaki[1,2] and Yoshinori Tokura[1,2]†

[1]*RIKEN Center for Emergent Matter Science (CEMS), Wako 351-0198, Japan*

[2]*Department of Applied Physics and Quantum Phase Electronics Center (QPEC), University of Tokyo, Tokyo 113-8656, Japan*

[3]*PRESTO, Japan Science and Technology Agency (JST), Chiyoda-ku, Tokyo 102-0075, Japan.*

[4]*Institute for Materials Research, Tohoku University, Sendai 980-8577, Japan*

†To whom correspondence should be addressed (okada@cmr.t.u-tokyo.ac.jp, tokura@riken.jp)

*These authors equally contributed to this work.

**Electrodynamic responses from three-dimensional (3D) topological insulators (TIs)[1,2] are characterized by the universal magnetoelectric $E \cdot B$ term constituent of the Lagrangian formalism[3,4]. The quantized magnetoelectric coupling, which is generally referred to as topological magnetoelectric (TME) effect[3,4], has been predicted to induce exotic phenomena[3,5-9] including the universal low-energy magneto-optical effects[3,6,7]. Here we report the experimental demonstration of the long-sought TME effect, which is exemplified by magneto-optical Faraday and Kerr rotations[3,6,7] in the quantum anomalous Hall (QAH) states of magnetic TI surfaces[10-15] by terahertz magneto-optics. The universal relation composed of the observed Faraday and Kerr rotation angles but not of any material parameters (e.g. dielectric constant and magnetic susceptibility) well exhibits the trajectory toward the fine structure constant $\alpha$ ($= 2\pi e^2/hc \sim 1/137$) in the quantized limit. Our result will pave a way for versatile TME effects with emergent topological functions[3,5,8,9].**

Topological quantum phenomena have been attracting increasing attention in condensed matter physics, because the system determined by the topological structure exhibits quantized observables, such as magnetic flux in superconductors and Hall conductance in quantum Hall effect. Magnetoelectric coupling, which has been a fundamental concept for contemporary physics including spintronics and multiferroics[16], is predicted to be quantized in the recently discovered 3D TIs[1-4]. More specifically, quantized magnetoelectric responses are predicted upon the QAH state induced by the magnetic mass-gap on the surface Dirac cone under the broken time-reversal symmetry. In the QAH state, as experimentally confirmed recently[10-15], the surface states exhibit quantized Hall conductance ($\sigma_{xy} = e^2/h$ and $\sigma_{xx} = 0$) without external magnetic field.

The electromagnetic field in 3D TIs is characterized by the Lagrangian including a novel magnetoelectric term, as generally referred to as the axion term[17], leading to the TME effect[3,4]. The axion term $\left(\frac{\alpha}{4\pi^2}\right)\theta \boldsymbol{E}\cdot\boldsymbol{B}$ ($\alpha$ being the fine structure constant and $\theta$ the $Z_2$ topological number: $\theta = \pi$ for TIs and $\theta = 0$ for trivial insulators) endows the TI with magnetoelectric susceptibilities quantized in units of $\alpha$. A promising manifestation of TME effect is magneto-optics[3,6,7], where the rotation angle of polarization is essential quantity. The optical responses of QAH state of a magnetic TI exhibit quantized Faraday and Kerr rotation angles[3,6,7], which represent polarization rotations for transmission and reflection geometries, respectively. In fact, the relation denoted with rotation angles ($\theta_F$ and $\theta_K$) always leads to the fine structure constant $\alpha = 2\pi e^2/hc = 1/137$, irrespective of material parameters such as dielectric constant and magnetic susceptibility, whereas the magneto-optical rotation angles of a thin film on substrate are substantially modified from those of the free-standing film in vacuum ($\theta_F = \alpha \sim 7.3$ mrad and $\theta_K = \pi/2$ rad)[3,6,7]. Thus the observation of Faraday and Kerr rotations on QAH state provides a measure of $\alpha$. However, the difficulties in experimental verification of TME effect have been indicated since the early stage of theoretical predictions[3,6,7]. This is because the observation of TME effect requires the Fermi energy within the magnetic mass-gap on the surface Dirac cone, and hence precise Fermi energy tuning is indispensable. In addition, the observation of genuine TME signal is limited to the low energy, i.e., sufficiently lower than the magnetic mass-gap to avoid the responses from real electronic transitions.

The QAH state on the surfaces of TI is stabilized by the magnetic mass-gap of about 50 meV for Cr: $(Bi, Sb)_2Te_3$ (ref. 18), so that the magneto-optics by terahertz (THz) spectroscopy (< 10 meV) can be applied to the observation of the possible emergence of TME effect. Furthermore, the recently developed magnetic modulation-doping in Cr:(Bi, Sb)$_2$Te$_3$ thin film[14] can dramatically widen the observable temperature region of QAH effect up to several Kelvin, making feasible the optical measurement of QAH state. So far the low-energy magneto-optical responses have been intensively studied[19-26] mostly for nonmagnetic TIs, in which the cyclotron resonances of the surface states as well as the bulk carriers are reported. However, an experimental demonstration of TME effect on QAH state remains elusive. In this paper, we show the first experimental observation of TME effect in QAH states on a magnetic TI thin film by THz time-domain spectroscopy (TDS). The trajectory towards the fine structure constant $\alpha$ is unveiled by the measurements of THz Faraday and Kerr rotation angles for the surface QAH state.

The Cr$_x$(Bi$_{0.26}$Sb$_{0.74}$)$_{2-x}$Te$_3$ TI film with magnetic modulation-doping[14], where magnetic impurities Cr ($x = 0.57$) are doped in two quintuple layers adjacent to the top and bottom layers, is schematically illustrated in Fig. 1a. The evolution of the magnetization induced by the Cr-doping gives rise to the QAH state as shown in Figs. 1b and 1c. The ferromagnetic transition occurs around $T_C \sim 70$ K with the onset of the anomalous Hall term in $\sigma_{xy}$ (Fig. 1b). As temperature decreases, $\sigma_{xy}$ develops and tends to saturate at the quantized value $e^2/h$ at around $T = 0.5$ K, while $\sigma_{xx}$ steeply decreases towards zero, due to the emergence of the dissipationless chiral edge conduction (see Fig. 1b). The Hall angle ($\sigma_{xy}/\sigma_{xx}$) becomes as large as 1 around 4 K, indicating the emergence of QAH regime at temperatures more than an order of magnitude higher[14] than the uniformly Cr-doped TI films[10-13,15], due perhaps to the enlargement of the magnetic mass-gap induced by the rich Cr-doping and the reduced disorder in the surface states by Cr dopants[14]. Hysteretic behaviors of Hall conductance further evidence the development of the QAH regime as shown in Fig. 1c. The fully quantized $\sigma_{xy}$ at the lowest temperature indicates that the Fermi energy locates well within the magnetic mass-gap of the surface Dirac cone (Fig. 1a) without additional field-effect tuning.

THz-TDS provides magneto-optical measurements with sufficiently lower photon energy (1 ~ 8 meV) than the magnetic mass-gap (~ 50 meV)[18] and with high resolution of light-polarization rotations (< 1 mrad). The measurement configuration of magneto-optics by THz-TDS is schematically illustrated in Fig. 1d (see Methods for detail). Depending on the time delay, the monocycle THz pulse can differentiate the directly transmitted pulse (i) and the delayed pulse generated by back-reflection at the back surface of substrate (ii), as shown in Figs. 1d and 1e; this enables us to separably measure Faraday and Kerr rotations, as reported for TI thin films[19,21] and graphene on substrates[27]. As shown in Fig. 1e, the temporal waveform of $E_y$-component indicates the pronounced rotation of polarization on the first pulse (i) as well as on the second one (ii) due to the presence of the magneto-optical rotations at zero external magnetic field. The first pulse (i) involves the Faraday rotation ($\theta_F$), while the second pulse (ii) is composed of $\theta_F$ plus the Kerr rotation ($\theta_K$) at the back-surface of the magnetic TI film (Fig. 1d).

The transmittance spectra obtained by $E_x$-component at different temperatures are shown in Fig. 2a. The transmittance is close to unity, i.e. no absorption, except for the dips around 7 meV indicated by the arrow, which is assigned to the optical phonon mode[28]; see also the optical conductance $\sigma_{xx}$ spectrum at $T$ = 4.3 K in Fig. 2a. The negligibly weak absorption, e.g. no Drude response, confirms that the Fermi energy locates within the magnetic mass-gap of the surface Dirac cone (Fig. 1a).

Fourier transformation of the electric field pulses $E_x(t)$ and $E_y(t)$ (Fig. 1e) provides the complex Faraday and Kerr rotation spectra (Figs. 2b and 2c), where the real part, $\theta_F(\omega)$ or $\theta_K(\omega)$, and the imaginary part, $\eta_F(\omega)$ or $\eta_K(\omega)$, represent the rotation angle and the ellipticity, respectively (see Methods for detail). The rotation-angle (real part) spectra for $\theta_F$ and $\theta_K$ show finite values around 2 and 6 mrad, respectively, with modest frequency dependence, as shown in Figs. 2b and 2c. The ellipticity (imaginary part) spectra for $\eta_F$ and $\eta_K$ are close to zero in the whole photon-energy region (< 8 meV). These characters, *i.e.* little frequency dependence and near-zero ellipticity, strongly indicate that the current THz energy window (1 ~ 8 meV) is well below the threshold energy for any magneto-optically active real transitions. This is consistent with the fact that the magnetic mass-gap on the Dirac point (~ 50 meV) is sufficiently large as compared to the energy range of this measurement. Note also that possible cyclotron resonance under magnetic field, which has been observed in previous magneto-optical studies on TIs[19-25], is absent in the present measurement because of zero external magnetic field. Furthermore, the observed Faraday and Kerr rotation angles are quantitatively consistent with the estimated rotation angles at dc limit (Figs. 2b and 2c), which are calculated from $\sigma_{xx}$ and $\sigma_{xy}$ obtained by the dc transport measurement (Fig. 1c) through the following equations[23,27];

$$\theta_F = \frac{1}{2}\arg\left[\frac{t_-}{t_+}\right] \quad \left(t_\pm = \frac{2}{1+n_s+Y_\pm}\right) \quad (1)$$

$$\theta_K = \frac{1}{2}\arg\left[\frac{r_-}{r_+}\right] \quad \left(r_\pm = \frac{-1+n_s-Y_\pm}{1+n_s+Y_\pm}\right). \quad (2)$$

Here, $t_{+(-)}$ and $r_{+(-)}$ represent the transmittance and reflection coefficients of right- and left-handed circularly-polarized light, respectively. The admittance $Y_\pm$ are described as $Y_\pm = Z_0(\sigma_{xx} \pm i\sigma_{xy})$ ($Z_0 = 377$ Ω: the vacuum impedance) and $n_s$ (~ 3.4) is the refractive index of the InP substrate[29]. For instance, the estimated rotation angles for $\theta_F$ and $\theta_K$ at $\omega = 0$ are around 3 and 9 mrad at 1.5 K (indicated with closed squares on the respective ordinates in Figs. 2b and 2c). This quantitative agreement with the dc QAH state exemplifies that the observed THz rotation stems from the TME effect on the TI surfaces. Figure 2d shows the temperature evolution of the Faraday and Kerr rotation spectra. The rotation angles decrease with increasing temperature and vanish at $T_C$ (see also Fig. 3a), in accord with the disappearance of the ferromagnetic state.

In Fig. 3b the rotation angles at different temperatures, which are measured by averaging the rotation angle below ~ 4 meV, are displayed as a function of the dc Hall conductance together with the calculated values from Eqs. (1) and (2) at $\omega = 0$. The observed THz Faraday and Kerr rotation angles show a good agreement with the estimated $\omega = 0$ value, although small deviations are discerned. The small reduction of the rotation angle from the dc limit, whose origin is probably relevant to some residual in-gap states[30] but not clear at present, has been also reported in THz magneto-optics of quantum Hall effect in graphene[27].

The relationship between the Faraday and Kerr rotation angles at the quantized limit is expected to lead to the fine structure constant $\alpha$, irrespective of any material parameters such as the dielectric constant and the thickness of the film, the capping layer and the substrate[6]. In our measurement geometry, the universal relationship between $\theta_F$ and $\theta_K$ in the QAH state is obtained from Eqs. (1) and (2), following the procedure proposed in ref. 6;

$$\frac{\cot\theta_F - \cot\theta_K}{\cot^2\theta_F - 2\cot\theta_F \cot\theta_K - 1} = \alpha. \quad (3)$$

Here we define the left side of Eq. (3) as the scaling function $f(\theta_F, \theta_K)$. In Fig. 3c,

the function $f(\theta_F, \theta_K)$ versus dc Hall conductance $\sigma_{xy}^{dc}$ is plotted, in which the $f(\theta_F, \theta_K)$ is expected to reach $\alpha$ (= $2\pi e^2/hc \sim 1/137$) in the quantized limit. With increasing $\sigma_{xy}$ to the quantized conductance by lowering temperature, the dimensionless $f(\theta_F, \theta_K)$ approaches the universal value $\alpha$, in good agreement with the estimation at $\omega = 0$ (line in Fig. 3c), manifesting the trajectory toward the quantized value $\alpha$ determined uniquely and solely by the magneto-optical rotation angles.

In conclusion, we have experimentally investigated the TME effect on the QAH state of surface state of TI by measurements of Faraday and Kerr rotations in THz region. The observed Faraday and Kerr rotation angles show quantitative agreement with the estimation from the dc transport results. The universal relationship with the magneto-optical rotation angles shows the trajectory converging to the fine structure constant $\alpha$ with the approach to the QAH state.

**Methods**

**Sample fabrication.** The 8-nm-thick TI films with magnetic modulation-doping were grown on both-side-polished insulating InP substrates by molecular beam epitaxial growth as described in ref. 14. To protect the film from degradation, a 3-nm-thick AlO$_x$ layer was immediately deposited with *ex-situ* atomic-layer-deposition. Transport measurement and optical THz spectroscopy were performed on different samples from the same batch. Possible modification of rotation angles by the AlO$_x$ capping layer is estimated to be as small as 0.01 % at most, and hence neglected in the analysis described in the main text.

**Magneto-optical terahertz spectroscopy.** In the THz-TDS, laser pulses with duration of 100 fs from a mode-locked Ti: sapphire laser were split into two paths to generate and detect THz pulses. THz pulses were generated by a bow-tie shaped antenna and detected by a dipole antenna. The $E_y(t)$ component of the transmitted THz pulses (Fig. 1e) was measured by the Crossed-Nicole configuration by using wire-grid polarizers. The polarization rotation $E_y(t)$ at 0 T is defined by the antisymmetrized waveform to eliminate the background signal, which is the difference between signals with magnetization for $\pm z$ directions after the poling of the magnetization at $\pm$ 1 T; $E_y(t) = (E_y^{+M}(t) - E_y^{-M}(t))/2$. The Fourier transformation of the first THz pulses $E_x(t)$ and $E_y(t)$ (Fig. 1e) gives the complex Faraday rotation spectra $E_y(\omega)/E_x(\omega) = (\sin\theta_F(\omega) + i\eta_F(\omega)\cos\theta_F(\omega))/(\cos\theta_F(\omega) - i\eta_F(\omega)\sin\theta_F(\omega)) \sim \theta_F(\omega) + i\eta_F(\omega)$ (Fig. 1b) for the small rotation angles. The rotation spectra obtained by the second pulses give the sum of the Kerr and Faraday rotation spectra. Transmittance spectra were obtained by comparison between the transmission of sample and bare substrate. We applied the following standard formula to obtain the complex conductance $\sigma(\omega) = \sigma_1(\omega) + i\sigma_2(\omega)$ of TI film;

$$t(\omega) = \frac{1+n_s}{1+n_s+Z_0\sigma(\omega)},$$

where $t(\omega)$ is the complex transmittance, $Z_0$ is the impedance of free space (377 Ω) and $n_s$ the refractive index of the InP substrate.

**Acknowledgements**

This research was supported by the Japan Society for the Promotion of Science through the Funding Program for World-Leading Innovative R & D on Science and Technology (FIRST Program) on "Quantum Science on Strong Correlation" initiated by the Council for Science and Technology Policy and by JSPS Grant-in-Aid for Scientific Research(S) No. 24224009, 24226002 and 26706011. K.N.O. is supported by RIKEN Junior Research Associate Program.

**Author contributions**

Y. Tokura conceived the project.   K.N.O. and Y. Takahashi carried out optical terahertz spectroscopy and analyzed data.   M.M., R.Y., K.N.O., A.T., and K.S.T. prepared the modulation-doped topological insulator thin films and performed the structural and transport characterizations.   The results were discussed and interpreted by K.N.O., Y. Takahashi, N.O., M.K., A.T. and Y. Tokura.

**Additional information**

The authors declare no competing financial interests.

**Figure 1| Terahertz (THz) Faraday and Kerr rotation of quantum anomalous Hall (QAH) state on magnetic topological insulator (TI) film.   a.** Schematics of the TI film with magnetic modulation doping and of the band structure of surface states under the presence of time-reversal-symmetry breaking magnetization.   **b.** Temperature dependence of the longitudinal ($\sigma_{xx}$) and Hall ($\sigma_{xy}$) conductances at $B$ = 0.1 T.   **c.** Magnetic field dependence of $\sigma_{xy}$ at various temperatures.   **d.** Schematics of the THz magneto-optics for the magnetic TI film on an InP substrate.   Crossed-Nicol geometry was employed for the detection of the magneto-optical rotation of light polarization. Faraday and Kerr rotations are measured by the first THz pulse (i) and the second THz pulse (ii), respectively.   **e.** Time evolutions of THz pulses through the magnetic TI film at 0 T after the poling of magnetic moment. (see Methods for detail.)   $E_x$ and $E_y$ are transmitted light polarized parallel and perpendicular to the incident light, respectively. The first pulse (i) represents the directly transmitted light through the TI film and substrate, while the THz pulse once reflected at back-surface of substrate appears as the second pulse (ii) with a time delay.   $E_y$ component of the second pulse includes the Kerr rotation $\theta_K$ at back-surface of TI film as well as the Faraday rotation $\theta_F$.

**Figure 2| THz Faraday and Kerr rotation spectra of the magnetic TI film. a.** Transmittance (left ordinate) spectra of the magnetic TI film at various temperatures and optical conductance (right ordinate) $\sigma_{xx}$ spectrum at $T$ = 4.3 K.   The transmittance close to unity was observed, indicating the negligible carrier absorptions due to fine-tuning of the Fermi level within the magnetic mass-gap of the surface state.   The low-lying optical phonon mode is discerned as a tiny dip around 7 meV and also as the

peak of optical conductance spectrum. **b, c.** Complex Faraday (**b**) and Kerr (**c**) rotation spectra at 1.5 K (see the main text and Method for definition). The real parts ($\theta_F$ and $\theta_K$) represent the rotation angle of light polarization. The imaginary parts ($\eta_F$ and $\eta_K$) represent the ellipticity, which is negligibly small as expected. Rotation angles at $\omega = 0$ evaluated from the dc transport data (Fig. 1c) are also plotted on the left ordinates. **d.** Temperature dependence of the Faraday (red) and Kerr (blue) rotation spectra with the evaluated $\omega = 0$ values on the ordinates.

**Figure 3| Trajectory toward the quantized topological magnetoelectric (TME) response.** **a.** Temperature dependence of the Faraday (red) and Kerr (blue) rotation angles, which are measured by averaging the rotation spectra below ~4 meV (see Fig. 2d). **b.** The observed Faraday (red) and Kerr (blue) rotation angles, $\theta_F$ and $\theta_K$, versus dc Hall conductance $\sigma_{xy}^{dc}$ (see Figs. 1c and 3a). The solid lines represent the estimation from Eqs. (1) and (2) in the main text. **c.** Evolution of the scaling function $f(\theta_F, \theta_K) = \dfrac{\cot\theta_F - \cot\theta_K}{\cot^2\theta_F - 2\cot\theta_F \cot\theta_K - 1}$ (see the main text and Eq. (3)) without any material parameters as a function of dc Hall conductance, which is expected to reach the fine structure constant $\alpha$ (~ 1/137) in the quantized limit, as indicated by a straight line.

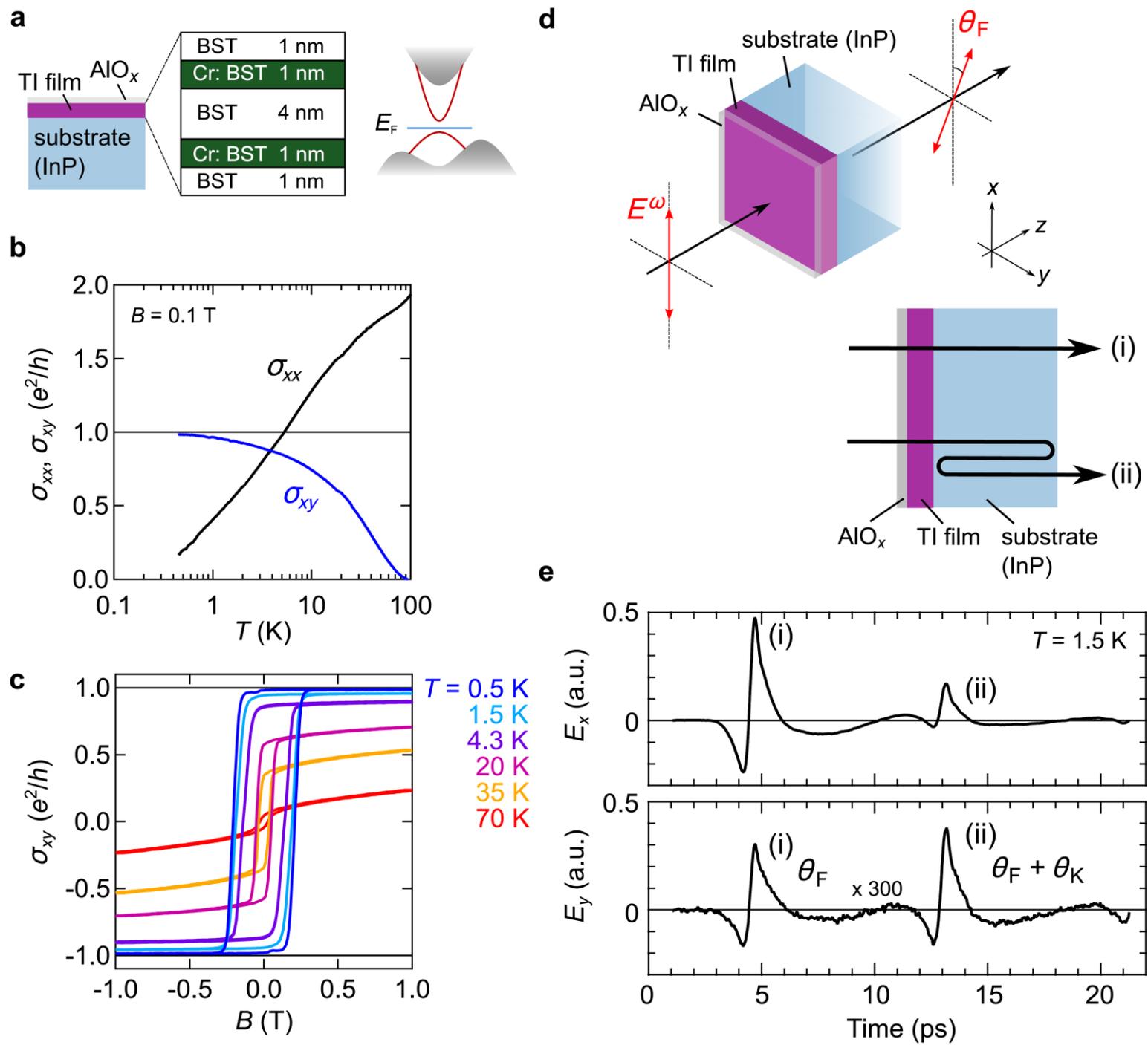

Figure 1   K. N. Okada and Y. Takahashi *et al*.

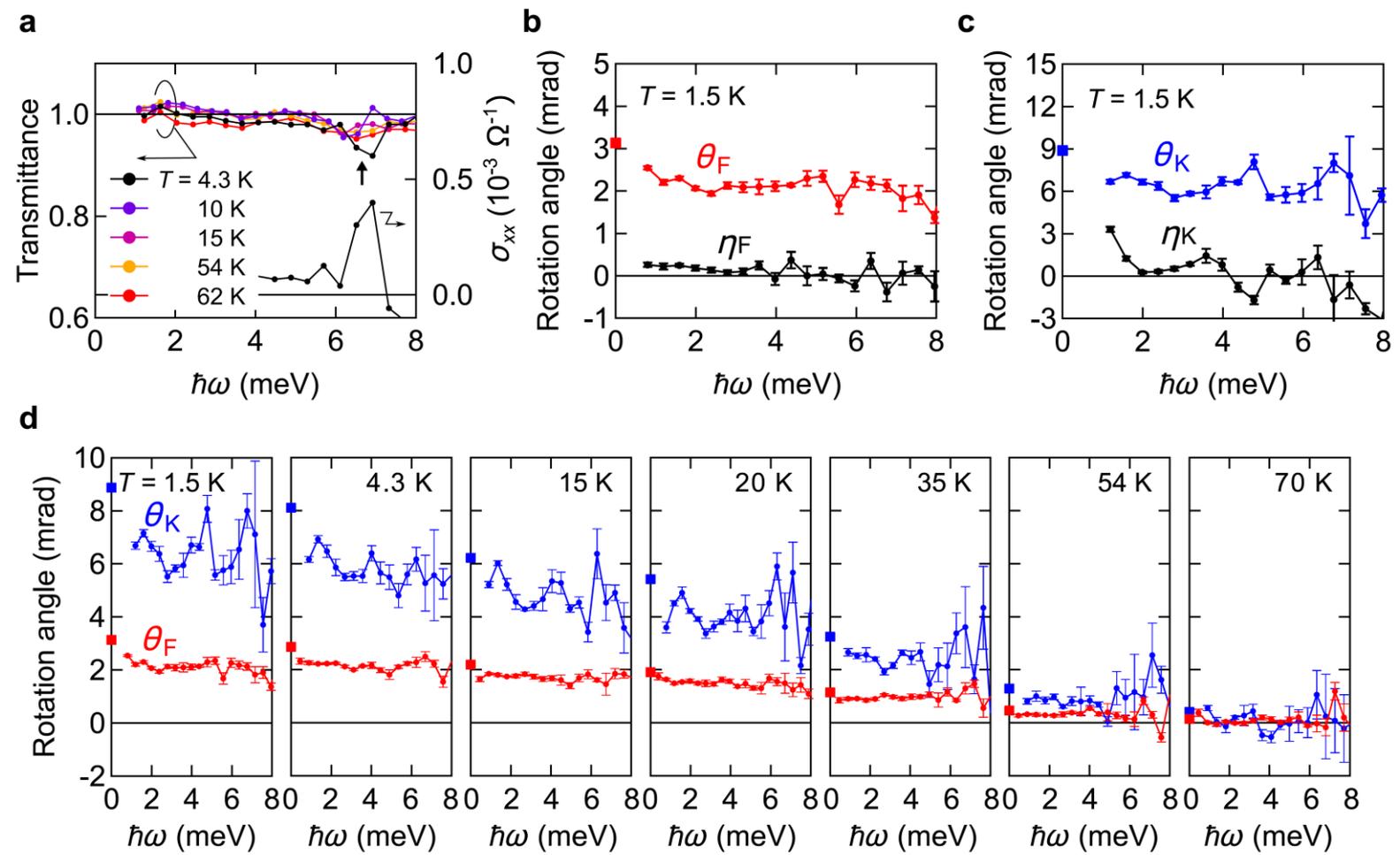

Figure 2  K. N. Okada and Y. Takahashi *et al*.

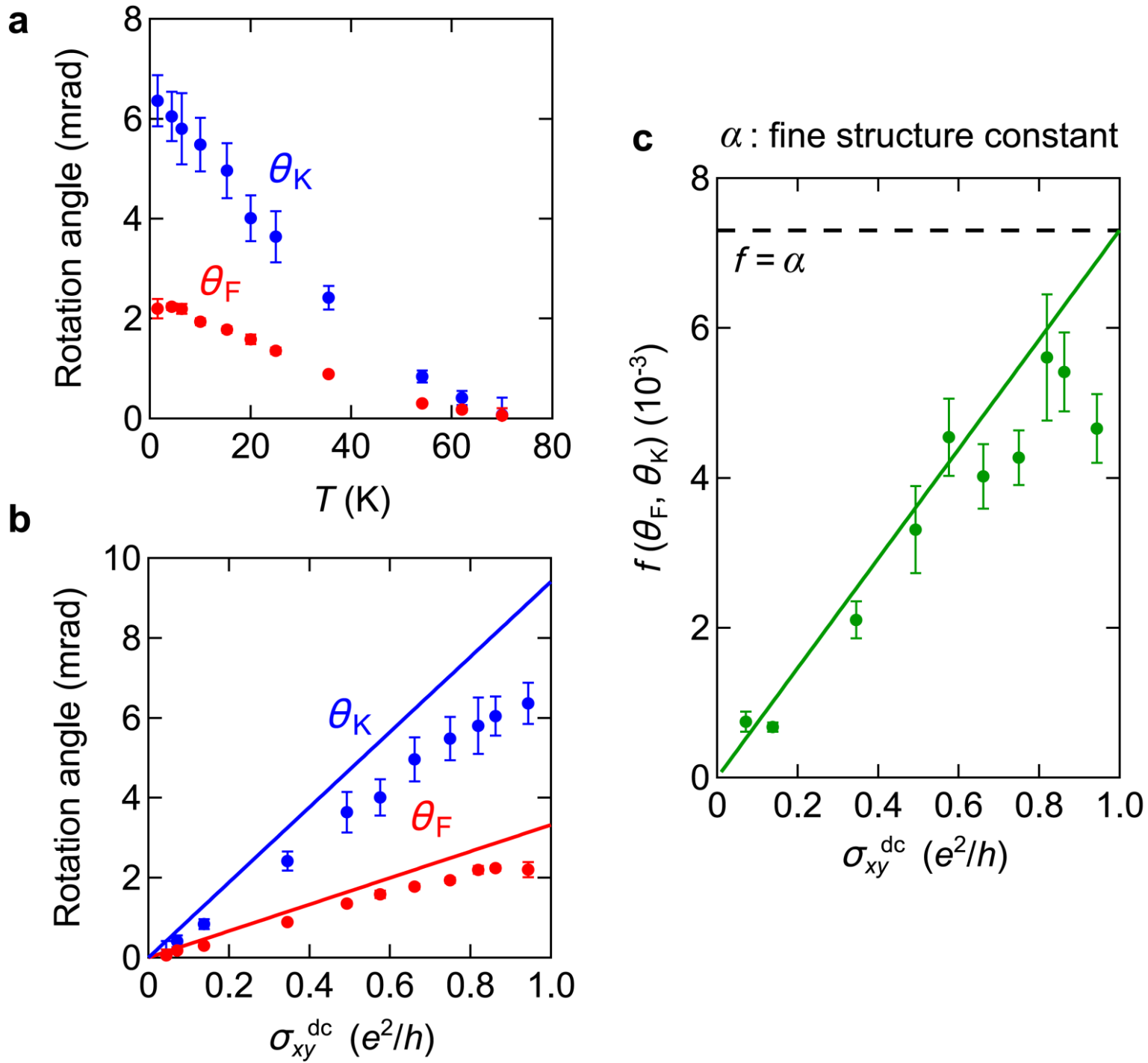

Figure 3    K. N. Okada and Y. Takahashi *et al*.